\documentclass[11pt]{article}
\usepackage{amsmath,amssymb,amsthm,latexsym}
\usepackage[dvips]{graphics}
\usepackage[english]{babel}
\usepackage{amsfonts}
\usepackage{array}
\usepackage{color}
\usepackage[colorlinks,citecolor=blue,urlcolor=blue]{hyperref}

\newtheorem{theorem}{{Theorem}}

\numberwithin{equation}{section} \numberwithin{theorem}{section}
\numberwithin{lemma}{section}\numberwithin{corollary}{section}

\usepackage
{graphics}
\usepackage
{graphicx}
\numberwithin{equation}{section}

\begin{document}

\noindent {\bf \Large Adjusted Empirical Likelihood Method for the Tail Index of A Heavy-Tailed Distribution}

\vspace{10pt}
\noindent{\bf Yizeng Li,~~~ Yongcheng Qi$^*$}

\vspace{10pt}

{
\vspace{10pt}

\noindent Department of Mathematics and Statistics, University of Minnesota Duluth,
1117 University Drive, Duluth, MN 55812, USA.
\vspace{5pt}

\noindent{$^*$\em Email address for corresponding author: yqi@d.umn.edu}

\date{\today}

\vspace{10pt}

\noindent{\bf Abstract.}
Empirical likelihood is a well-known nonparametric method in statistics and has been widely applied in statistical inference. The method has been employed by Lu and Peng (2002) to constructing confidence intervals for the tail index of a heavy-tailed distribution.
It is demonstrated in Lu and Peng (2002) that the empirical likelihood-based confidence intervals performs better than confidence intervals based on normal approximation in terms of the coverage probability.  In general, the empirical likelihood method can be hindered by its imprecision in the coverage probability when the sample size is small. This may cause a serious undercoverage issue when we apply the empirical likelihood to the tail index as only a very small portion of observations can be used in the estimation of the tail index.  In this paper, we employ an adjusted empirical likelihood method, developed by Chen et al. (2008) and Liu and Chen (2010), to constructing confidence intervals of
the tail index so as to achieve a better accuracy.
We conduct a simulation study to compare the performance of the adjusted empirical likelihood method and the normal approximation method. Our simulation results indicate that the adjusted empirical likelihood method outperforms other methods in terms of the coverage probability and length of confidence intervals. We also apply the adjusted empirical likelihood method to a real data set.

\vspace{20pt}

}


\newpage

\section{Introduction}
\label{intro}

In last few decades, heavy-tailed distributions have found applications in many fields such as meteorology, hydrology, climatology, environmental science, telecommunications, insurance and finance. The regularity of these distributions can be conveniently used to
 estimate extreme tail probabilities and high quantiles by extrapolating from intermediate order statistics.
See, e.g., Hall and Weissman (1997), Danielsson et al. (1998), Danielsson and de Vries (1997), and Embrechts et al. (1998, 1999).

The problem of estimating the tail index of a heavy-tailed distribution has attracted much attention. Several estimators have been proposed in the literature; see, for example, Hill (1975), Pickands (1975), Hall (1982), Drees (1995), and de Haan and Peng (1998).  Among these estimators, Hill's estimator is well investigated in the literature.

In this paper, we are interested in constructing confidence intervals for the tail index. A straightforward approach to obtain a confidence interval is based on the normal approximation to some point estimators of the tail index such as Hill's estimator.
Lu and Peng (2002) employed empirical likelihood method to constructing confidence intervals for the tail index and
compared the performance of empirical likelihood method and several other methods.

The empirical likelihood was introduced by Owen (1988, 1990) for the mean vector of independent and identically distributed (i.i.d.) observations. In the last two decades, it has been extended to a wide range of applications. The empirical likelihood method produces confidence regions whose shape and orientation are determined entirely by the data, and it possesses some advantages over other methods like the normal approximation method. When the sample size is small, the empirical likelihood method may have a serious undercoverage problem.
 One reason is that the empirical likelihood function is not well defined as the true value of the mean is not an interior point of the convex hull of the data with a significant probability, see, e.g., Tsao (2004). To solve the problem,  Chen et al. (2008) proposed an adjusted empirical likelihood method by adding a pseudo-observation into the data set such that the new empirical likelihood ratio test statistic is always defined for any sample size.  Furthermore, Liu and Chen (2010) identified the optimal weight for this pseudo-observation so that the adjusted empirical likelihood achieves the high-order precision of the Bartlett correction.

In estimation of the tail index of a heavy-tailed distribution, only a few largest observations can be used.  Therefore, there is often a serious undercoverage problem when we apply the empirical likelihood method to inference of the tail index.  To solve this problem, in this paper, we employ the adjusted empirical likelihood method to make inference on the tail index and construct empirical likelihood-based confidence intervals.

The rest of the paper is organized as follows.
In Section~\ref{hillestimator}, we will introduce Hill's estimator for the tail index and confidence intervals based on the normal approximation of Hill's estimator.  In Section~\ref{ELM}, we present the empirical likelihood method and adjusted empirical likelihood method for the tail index and discuss how to choose the weight for the pseudo-observation so that the adjusted empirical likelihood method can achieve better precision.
In Section~\ref{simulation}, we conduct a simulation study to compare the performance of different methods in terms of the coverage probability and average interval length, and apply these methods to a real data set on Danish fire losses. Finally, in Section~\ref{proof}, we give a sketch of the proof of Wilks' theorem on the adjusted empirical likelihood ratio test statistic.


\section{Heavy-tailed distributions and Hill's estimator}\label{hillestimator}

Assume $X_1,\cdots,X_n$ are i.i.d. random variables with a cumulative distribution function $F$ satisfying
\begin{equation}\label{heavytail}
\lim\limits_{t\rightarrow\infty}\frac{1-F(tx)}{1-F(t)}=x^{-1/\gamma}
\end{equation}
for all $x>0$, where $\gamma>0$ is an unknown parameter and $1/\gamma$ is termed as the tail index of the distribution $F$. Without loss of generality, we assume that $F(0)=0$.


Let $X_{n,1}\le \cdots \le X_{n,n}$ denote the order statistics based on $X_1,\cdots,X_n$. The well-known Hill estimator for $\gamma$ is given by
\[
\hat{\gamma}_n=\frac{1}{k_n}\sum\limits_{i=1}^{k_n}\log(X_{n,n-i+1})-\log(X_{n,n-k_n}),
\]
where $k_n$ is a sequence of positive integers such that $k_n\rightarrow\infty$ and $\frac{k_n}{n}\rightarrow 0$ as $n\rightarrow\infty$; see, e.g., Hill (1975). Obviously, only a very small portion of observations are used to estimate the parameter $\gamma$.


In order to make an inference on $\gamma$, a condition stronger than \eqref{heavytail} is required. Throughout this paper, we assume that there exists a function $A(t)\rightarrow 0$ such that
\begin{equation}\label{secondorder}
\lim\limits_{t\rightarrow\infty}\frac{U(tx)/U(t)-x^{\gamma}}{A(t)}=x^{\gamma}\frac{x^{\rho}-1}{\rho}
\end{equation}
for all $x>0$, where $U(x)$ is the inverse function of $\frac{1}{1-F(x)}$, and $\rho\le 0$ is a second order parameter controlling the convergence rate in \eqref{heavytail}. It follows from  de Haan and Peng (1998) that
\begin{equation}\label{normalapp}
\sqrt{k_n}(\hat{\gamma}_n-\gamma)\xrightarrow{d}N(0,\gamma^2)
\end{equation}
if
\begin{equation}\label{kn}
k_n\rightarrow\infty,~~\frac{k_n}{n}\rightarrow 0~\textnormal{ and }~\sqrt{k_n}A(\frac{n}{k_n})\rightarrow0~\textnormal{ as }n\rightarrow\infty.
\end{equation}

Based on equation \eqref{normalapp},  a $100(1-\alpha)\%$ confidence interval for $\gamma$ is given by
\begin{equation}\label{IN}
I_{N}(1-\alpha)=\Big(\frac{\hat{\gamma}_n}{1+z_{\alpha/2}/\sqrt{k_n}},\frac{\hat{\gamma}_n}{1-z_{\alpha/2}/\sqrt{k_n}}\Big),
\end{equation}
where $z_{\alpha/2}$ denotes the upper $\alpha/2$-level critical value of the standard normal distribution.  Note that the confidence
interval $I_N(1-\alpha)$ is obtained by solving $\{\gamma: -z_{\alpha/2}/\sqrt{k_n}<(\hat\gamma_n-\gamma)/\gamma<z_{\alpha/2}/\sqrt{k_n}\}$.
A conventional confidence interval for $\gamma$ based on \eqref{normalapp} can also be defined as
\[
\{\gamma: -\frac{z_{\alpha/2}}{\sqrt{k_n}}<\frac{\hat\gamma_n-\gamma}{\hat\gamma_n}<\frac{z_{\alpha/2}}{\sqrt{k_n}}\}
=(\hat\gamma_n-\frac{z_{\alpha/2}\hat\gamma_n}{\sqrt{k_n}}, \hat\gamma_n+\frac{z_{\alpha/2}\hat\gamma_n}{\sqrt{k_n}}).
\]
Our simulation study indicates that this confidence interval has a very poor coverage probability when $k_n$ is small. Therefore, we will consider the confidence interval $I_N(1-\alpha)$ given in \eqref{IN}.

\section{Empirical likelihood and adjusted empirical likelihood}\label{ELM}

Define $y_i=i\big(\log(X_{n,n-i+1})-\log(X_{n,n-i})\big)$, $i=1,\cdots, k_n$. For any fixed integer $k\ge 2$, $y_1, \cdots, y_k$ are approximately i.i.d. random variables having an exponential distribution with mean $\gamma$; see e.g., Weissman (1978). Note that $k=k_n$ can change with $n$ if condition \eqref{kn} holds. In fact, since $\hat\gamma_n=\frac1{k_n}\sum^{k_n}_{j=1}y_j$, we can see from \eqref{normalapp} that
$y_1, \cdots, y_{k_n}$ behave asymptotically as if they were i.i.d random variables with mean $\gamma$. Therefore,
Lu and Peng (2002) applied  Owen's empirical likelihood method to the sample $y_1, \cdots, y_{k_n}$ and defined empirical likelihood ratio test statistic for the tail index $\gamma$ as
\[
l_{\textnormal{EL}}(\gamma)=-2\log\Big(\sup\Big\{\prod\limits_{i=1}^{k_n}(k_np_i): p_i\ge 0,~i=1,\cdots,k_n,\sum\limits_{i=1}^{k_n}p_i=1,\sum\limits_{i=1}^{k_n}p_iy_i(\gamma)=0\Big\}\Big),
\]
where $y_i(\gamma)=y_i-\gamma$ for $1\le i\le k_n$.
By the method of Lagrange multipliers, we have
\[
l_{\textnormal{EL}}(\gamma)=2\sum\limits_{i=1}^{k_n}\log(1+\lambda y_i(\gamma)),
\]
where $\lambda$ is determined by
\[
\sum\limits_{i=1}^{k_n}\frac{y_i(\gamma)}{1+\lambda y_i(\gamma)}=0.
\]
Lu and Peng (2002) proved under conditions \eqref{heavytail}, \eqref{secondorder} and \eqref{kn} that
\begin{equation}\label{ELchi}
l_{\textnormal{EL}}(\gamma_0)\xrightarrow{d}\chi^2_1,
\end{equation}
where $\gamma_0$ is the true value of $\gamma$.  Based on \eqref{ELchi}, an approximate $100(1-\alpha)\%$ confidence interval for $\gamma$ can be obtained as follows
\begin{equation}\label{ELi}
I_{\textnormal{EL}}(1-\alpha)=\{\gamma: l_{\textnormal{EL}}(\gamma)<\chi^2_{\alpha}\},
\end{equation}
 where $\chi^2_{\alpha}$ is the upper $\alpha$-level critical value of a $\chi^2_1$ distribution.

As we have noticed that only $k_n$ data points are allowed in the estimation, where $k_n$ satisfies condition \eqref{kn}. In general, the range of $k_n$ is
very limited. It is known that the bias of Hill's estimator $\hat{\gamma}_n$ will be non-ignorable when  $k_n$ is too large. For illustration, we take $F$ as
Fr\'echet($1$), which is defined in Section~\ref{simulation}.  One can verify that \eqref{secondorder} holds with $A(t)=(2t)^{-1}$, $\rho=-1$ and $\gamma=1$.
In this case, condition \eqref{kn} holds if and only if $k_n\to\infty$ and $k_n=o(n^{2/3})$.  If the sample size $n=1000$,   then $n^{2/3}$ is only
$100$. Therefore, we can only use the value of $k_n$ in a very small range for a practical inference solution, and the undercoverage problem  for the procedure above will naturally arise as we have discussed in Section~\ref{intro}.

To overcome the undercoverage problem and increase the accuracy of the chi-square approximation for the empirical likelihood ratio test statistics when $k_n$ is small,
we apply the adjusted empirical likelihood method developed by Chen et al. (2008) to the tail index. Define a pseudo-data point
\[
y_{k_n+1}(\gamma)=-\frac{a_n}{k_n}\sum^{k_n}_{i=1}y_i(\gamma)=-a_n(\hat\gamma_n-\gamma),
\]
where $a_n>0$.  With this extra data point,  an adjusted empirical likelihood ratio test statistic is defined as
\[
l_{\textnormal{AEL}}(\gamma)=-2\log\Big(\sup\Big\{\prod\limits_{i=1}^{k_n+1}((k_n+1)p_i): p_i\ge 0,~ i\ge 1,\sum\limits_{i=1}^{k_n+1}p_i=1,\sum\limits_{i=1}^{k_n+1}p_iy_i(\gamma)=0\Big\}\Big).
\]
By the method of Lagrange multipliers, we have
 \[
l_{\textnormal{AEL}}(\gamma)=2\sum\limits_{i=1}^{k_n+1}\log(1+\lambda y_i(\gamma)),
\]
where $\lambda$ is determined by
\[
\sum\limits_{i=1}^{k_n+1}\frac{y_i(\gamma)}{1+\lambda y_i(\gamma)}=0.
\]
The adjusted empirical likelihood by Chen et al. (2008) was originally proposed for the mean vector of i.i.d. observations.  One can
verify that there always exists a probability vector $(p_1, \cdots, p_{k_n+1})$ such that $\sum\limits_{i=1}^{k_n+1}p_iy_i(\gamma)=0$, and thus $l_{\textnormal{AEL}}(\gamma)$ is well defined. Although there are a large range of values for $a_n$ to take,  it is recommended in Chen et al. (2008)
that one take $a_n=\max(1, (\log k_n)/2)$ when $y_1, \cdots, y_{k_n}$ are i.i.d. random variables.

We have the following chi-square approximation theorem for empirical likelihood ratio test statistic  $l_{\textnormal{AEL}}(\gamma)$.

\begin{theorem}\label{thm1}
Assume conditions \eqref{heavytail}, \eqref{secondorder} and \eqref{kn} are held and $a_n=o(k_n^{2/3})$ as $n\to\infty$.
 Then
\begin{equation}\label{AELchi}
l_{\textnormal{AEL}}(\gamma_0)\xrightarrow{d}\chi^2_1,
\end{equation}
where $\gamma_0$ is the true value of $\gamma$.
\end{theorem}

Based on \eqref{AELchi}, an approximate $100(1-\alpha)\%$ confidence interval for $\gamma$ can be obtained as follows
\begin{equation}\label{iAEL}
I_{\textnormal{AEL}}(1-\alpha)=\{\gamma: l_{\textnormal{AEL}}(\gamma)<\chi^2_{\alpha}\},
\end{equation}
A further study in Liu and Chen (2010) reveals that the adjusted empirical likelihood for the mean of i.i.d. random variables is Bartlett
correctable if $a_n=b/2$, where
\begin{equation}\label{b}
b=\frac{\alpha_4}{2\alpha_2^2}-\frac{\alpha_3^2}{3\alpha_2^3},
\end{equation}
and $\alpha_r$ is the $r$-th central moment of the underlying distribution of the i.i.d. random variables.  For an exponential distribution
with mean $\gamma$, we have
\[
{\alpha}_r=r!\gamma^r\sum\limits_{j=0}^{r}\frac{(-1)^j}{j!},
\]
 and the constant $b$ defined in \eqref{b} is given by
 \[
 b=\frac{9\gamma^4}{2\gamma^4}-\frac{8\gamma^6}{3\gamma^6}=\frac{9}{2}-\frac{4}{3}=\frac{19}{6}.
 \]

In our paper, since $y_1, \cdots, y_{k_n}$ are approximately i.i.d. random variables with an exponential distribution with mean $\gamma$, we expect that the adjusted empirical likelihood ratio test statistic achieves a faster convergence rate to the chi-square distribution under conditions in Theorem~\ref{thm1} if we take $a_n=19/12$.

\section{Comparison study and a real data application}\label{simulation}
In this section,  we will carry out a simulation study so as to compare the performance of the methods introduced in Sections~\ref{hillestimator} and \ref{ELM}, and then apply these methods to analyze Danish fire loses data.

\subsection{Simulation study}

\begin{figure}[!htb]
\begin{center}
\includegraphics[scale=1]{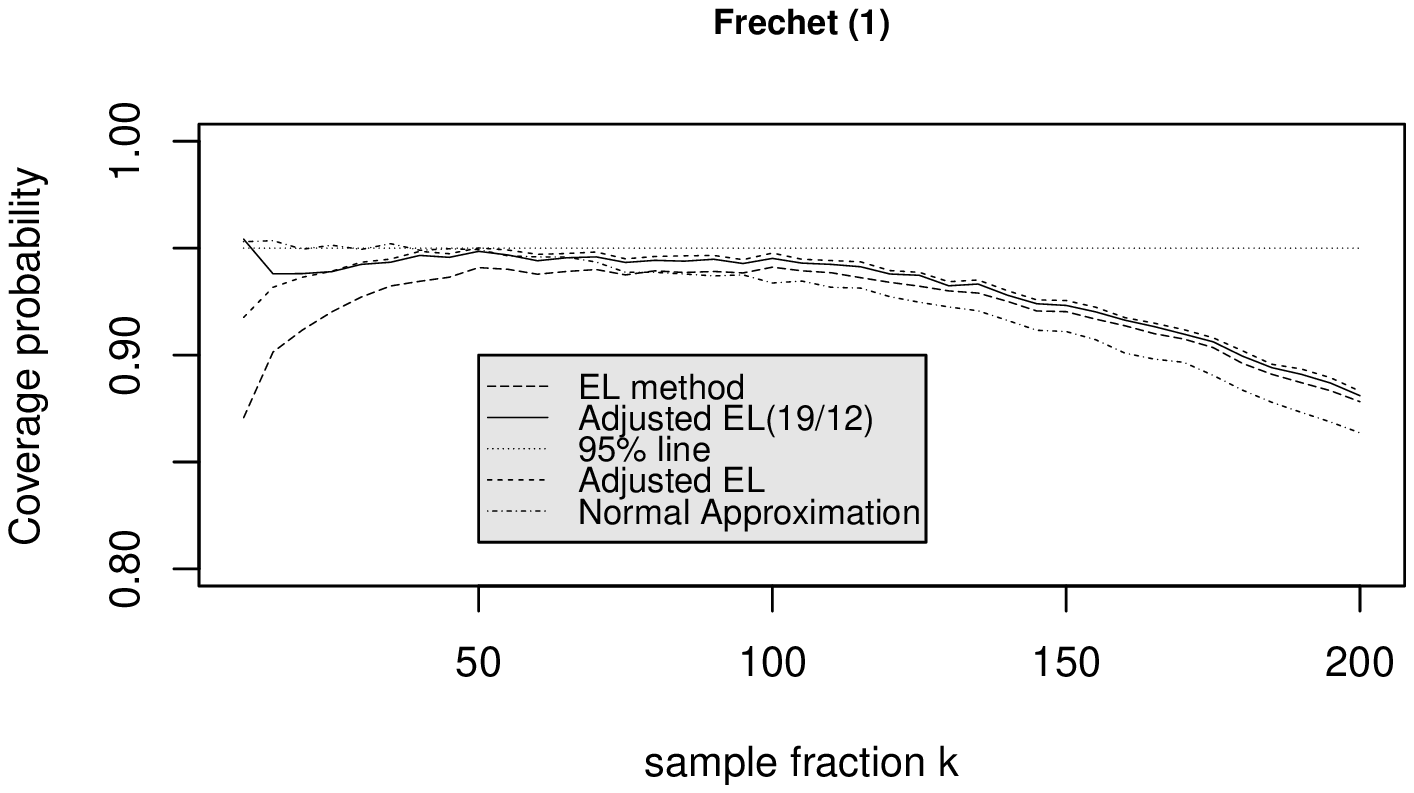}
\vspace{-20pt}
\includegraphics[scale=1]{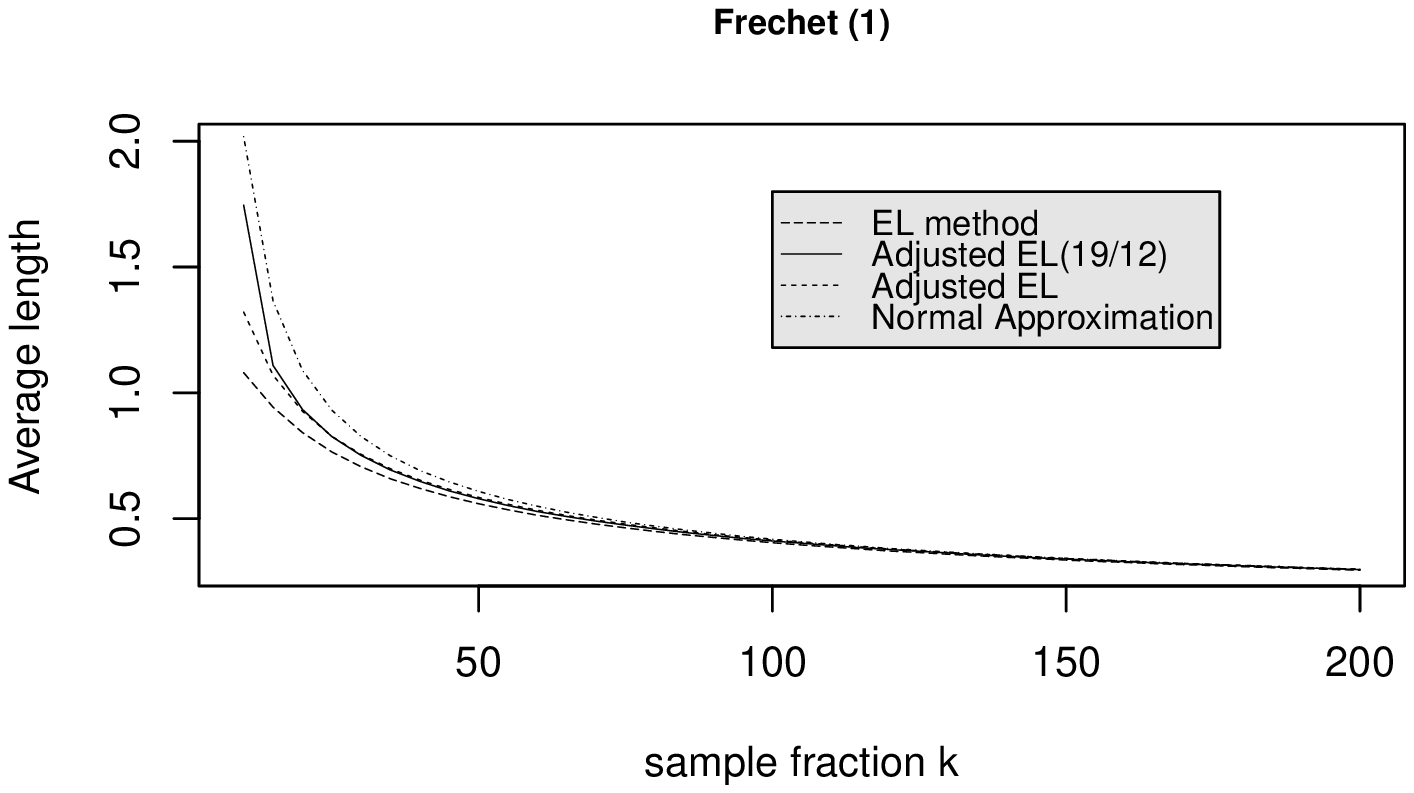}
\end{center}
\vspace{-25pt}
\caption{\sl Plot of the coverage probabilities and average lengths for confidence intervals from Fr\'echet($1$) distribution.  In the legend, ``Adjusted EL($19/12$)" stands for the adjusted empirical likelihood method with the correction factor $19/12$.}
\label{Figfrechet}
\end{figure}

\begin{figure}[!htb]
\begin{center}
\includegraphics[scale=1]{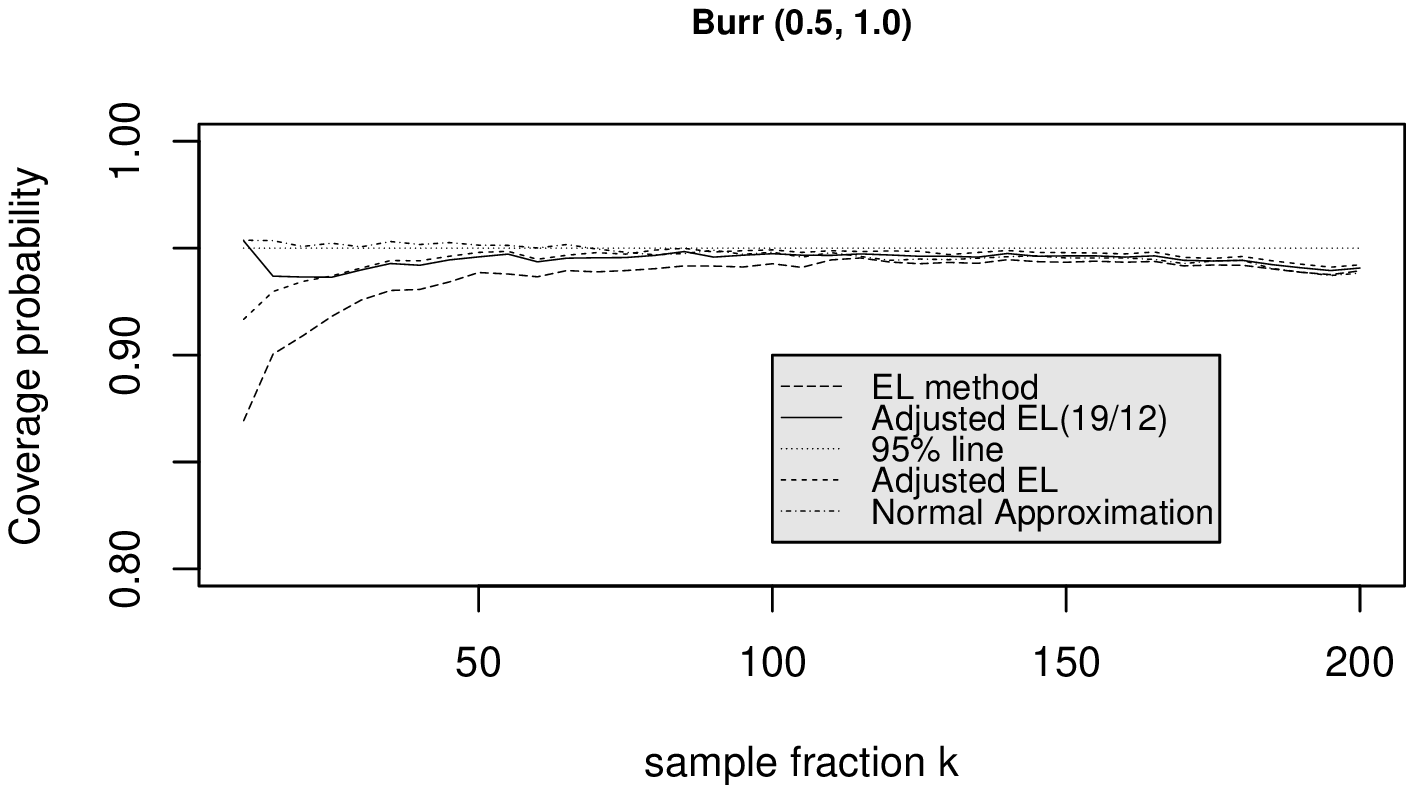}
\vspace{-20pt}
\includegraphics[scale=1]{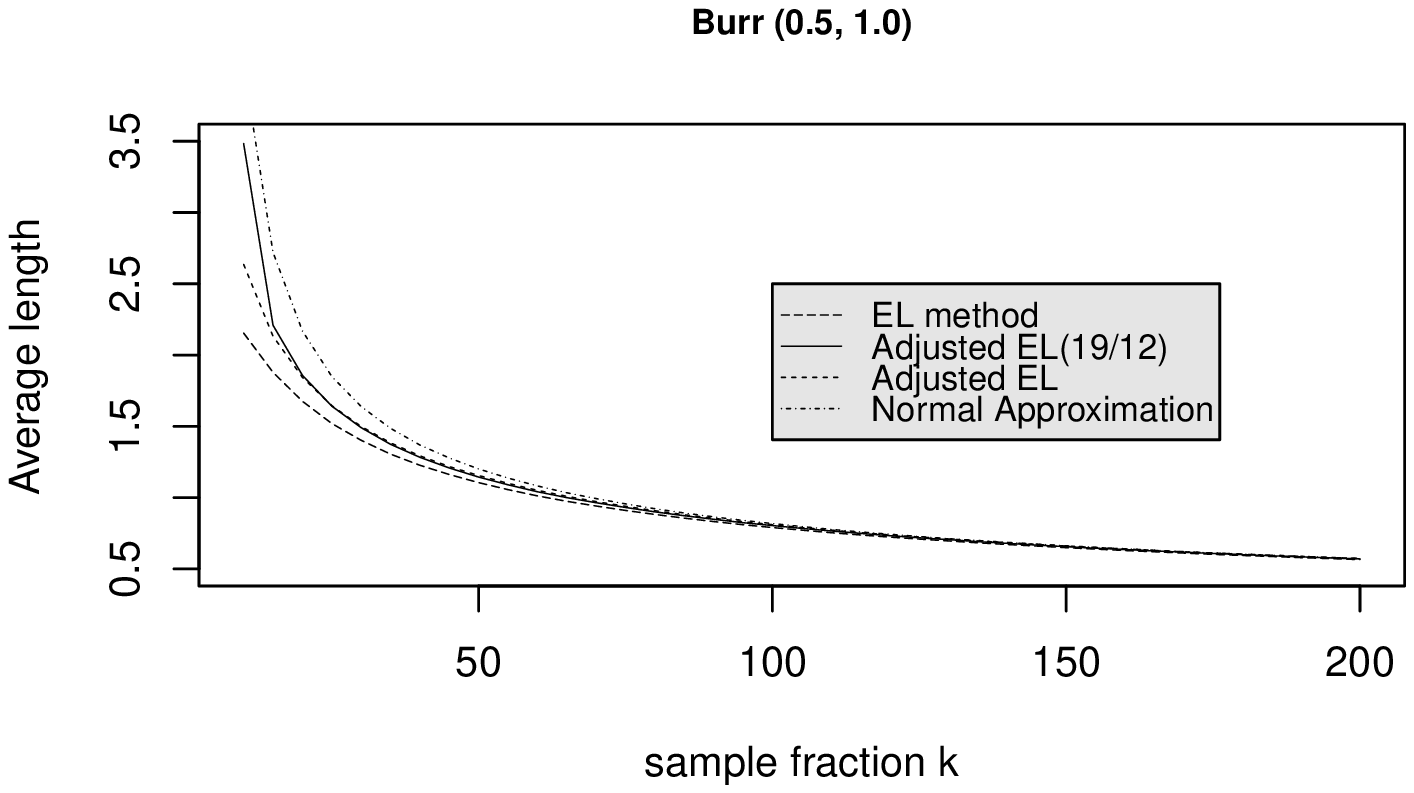}
\end{center}
\vspace{-25pt}
\caption{\sl Plot of the coverage probabilities and average lengths for confidence intervals from Burr($0.5$, $1.0$) distribution.  In the legend, ``Adjusted EL($19/12$)" stands for the adjusted empirical likelihood method with the correction factor $19/12$.}
\label{Figburr0510}
\end{figure}

\begin{figure}[!htb]
\begin{center}
\includegraphics[scale=1]{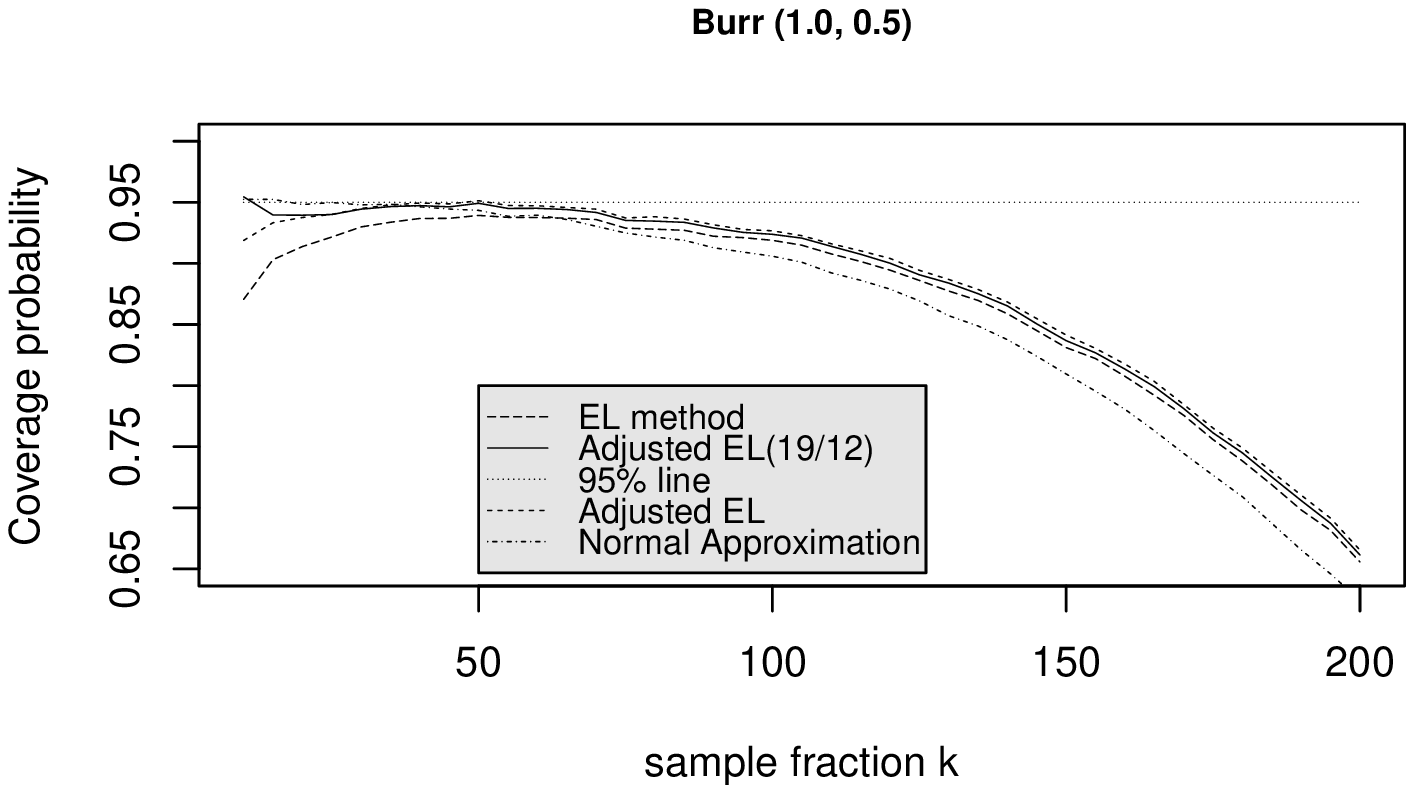}
\vspace{-20pt}
\includegraphics[scale=1]{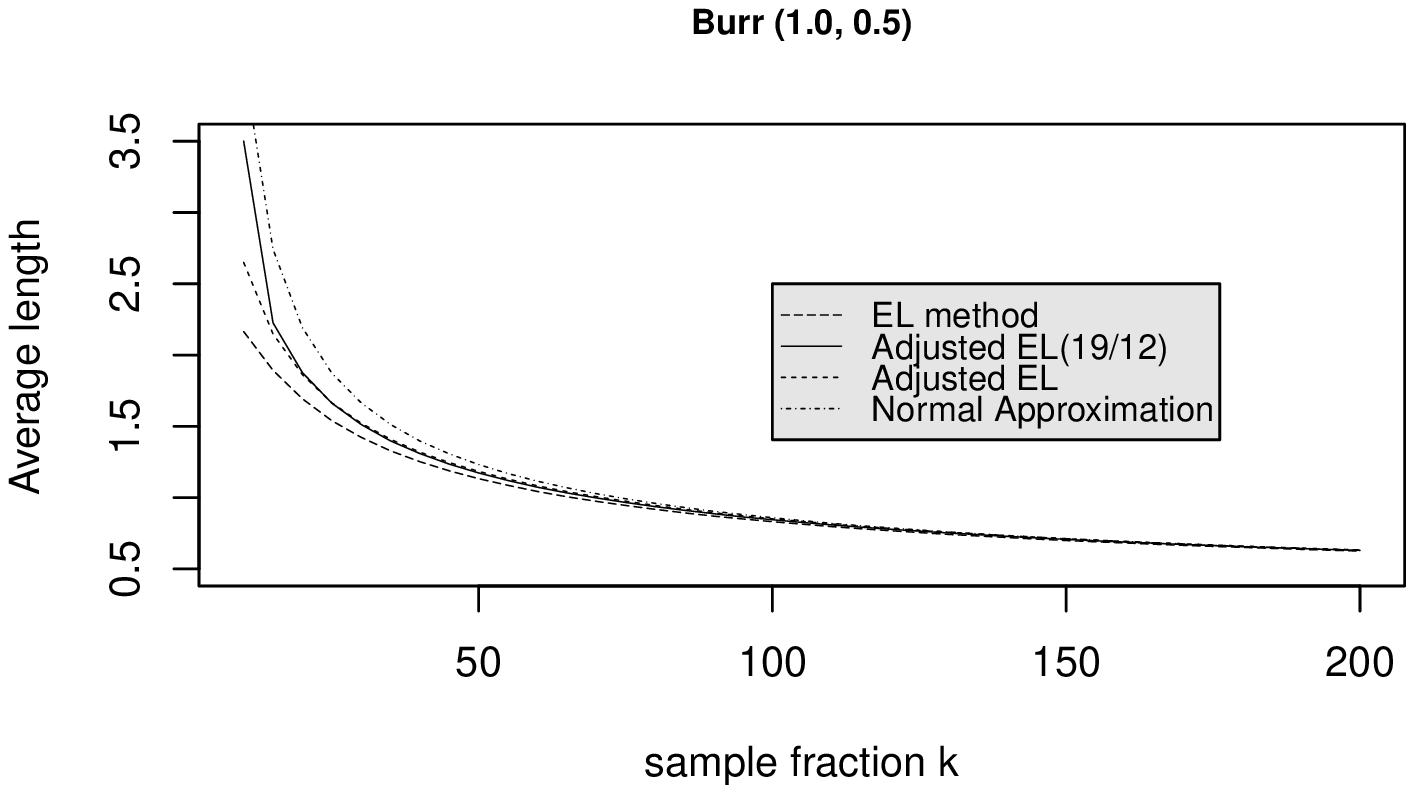}
\end{center}
\vspace{-25pt}
\caption{\sl Plot of the coverage probabilities and average lengths for confidence intervals from Burr($1.0$, $0.5$) distribution. In the legend, ``Adjusted EL($19/12$)" stands for the adjusted empirical likelihood method with the correction factor $19/12$.}
\label{Figburr1005}
\end{figure}

 We compare the five methods described above in terms of both the coverage probability and the average length of confidence intervals by using the following two types of heavy-tailed distributions.
\begin{itemize}
	\item[I.] The Fr\'echet distribution,  given by $F(x)=\exp(-x^{-\alpha})$ for all $x>0$, where $\alpha>0$  (notation Fr\'echet($\alpha$)). For this distribution, $\gamma=1/\alpha$.
	\item[II.] The Burr distribution, given by $F(x)=1-(1+x^{\alpha})^{-\beta}$ for all $x>0$, where $\alpha>0$ and $\beta>0$ (notation Burr($\alpha,\beta$)). We have $\gamma=1/(\alpha\beta)$.
\end{itemize}
In the simulation study, we drew $10\,000$ random samples of sample size $n=1000$ from Fr\'echet(1), Burr(0.5,1) and Burr(1,0.5) distributions, and then computed the coverage probabilities for $I_N(0.95)$, $I_{\textnormal{EL}}(0.95)$, $I_{\textnormal{AEL}}(0.95)$, $I_{\textnormal{AEL}^*}(0.95)$ (AEL with Bartlet correction factor) for $k=10,15,\cdots,200$;
see Figures~\ref{Figfrechet}, \ref{Figburr0510} and \ref{Figburr1005}. Confidence intervals $I_N(0.95)$, $I_{\textnormal{EL}}(0.95)$, $I_{\textnormal{AEL}}(0.95)$, and $I_{\textnormal{AEL}^*}(0.95)$ are defined in \eqref{IN}, \eqref{ELi} and \eqref{iAEL},  respectively.
Also we computed the average lengths of the intervals from all methods for $k=10,15,\cdots, 200$; see also Figures~\ref{Figfrechet}, \ref{Figburr0510} and \ref{Figburr1005}.

From the simulation results, we conclude that the adjusted empirical likelihood method with a correction factor $a_n=19/12$ outperforms other methods under consideration. First, the adjusted empirical likelihood method with the correction factor $19/12$ provides coverage probabilities which are comparable with those from the normal approximation for relatively small sample fractions $k_n$ and are constantly more accurate than the normal approximation when $k_n$ is getting larger; second, this method achieves shorter confidence intervals than the normal approximation method in general; third, this method has much better coverage probabilities than the other two empirical likelihood methods for small $k_n$ and has comparable performance for large $k_n$ in terms of the coverage probability and average length of confidence intervals.


\subsection{A real data application}\label{app}
\begin{figure}[!htb]
\begin{center}
\includegraphics[scale=1]{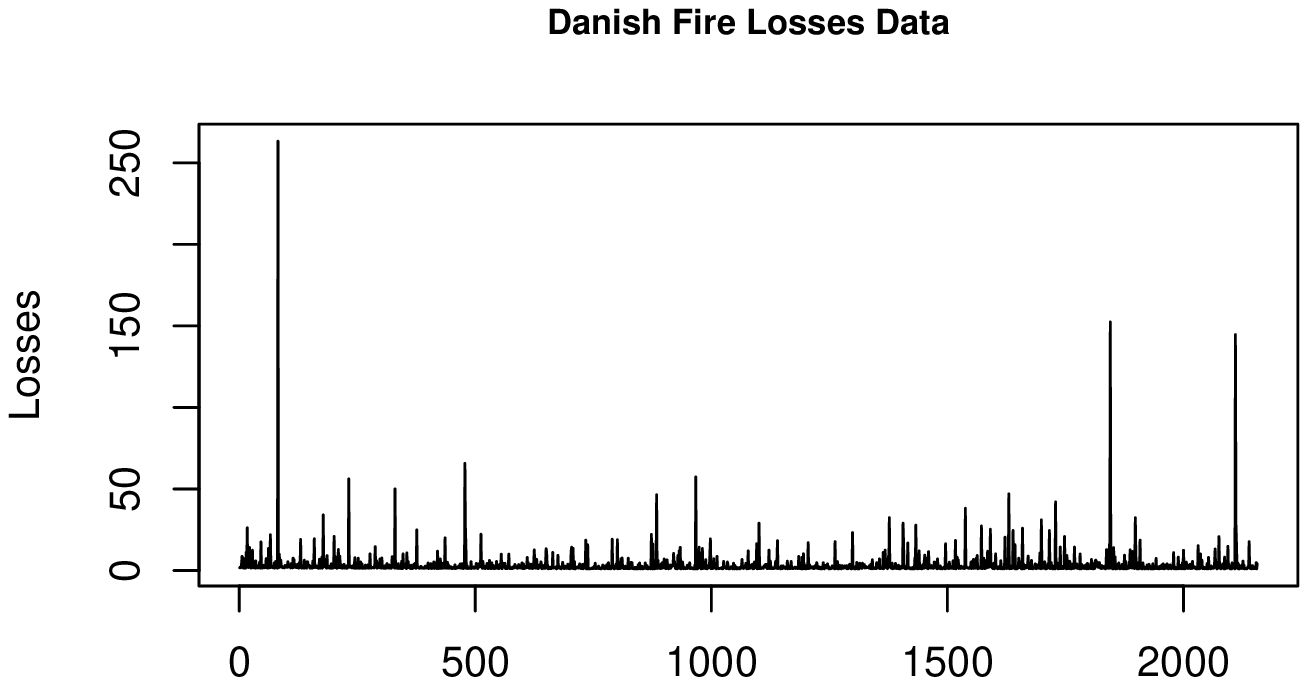}
\end{center}
\vspace{-25pt}
\caption{\sl  Danish Fire Losses Data}
\label{fig1}
\end{figure}

As an application, we apply empirical likelihood methods  and normal approximation methods to Danish fire dataset.
The data set contains $2156$ Danish fire losses over one
million Danish krone from year $1980$ to year $1990$ inclusive. These are total losses for events, including damage to
buildings, furnishings and personal property as well as loss of profits.
This dataset has been embedded in the \textbf{R} package ``QRM'' and it is also available at
 \url{www.ma.hw.ac.uk/~mcneil/.}
Fig.~\ref{fig1}
displays the data. This Danish fire losses dataset has been analyzed by McNeil (1997) and Resnick (1996).  See also Lu and Peng (2002) and Peng and Qi (2017) for comparison study.

\begin{figure}[!htb]
\begin{center}
\includegraphics[scale=0.85]{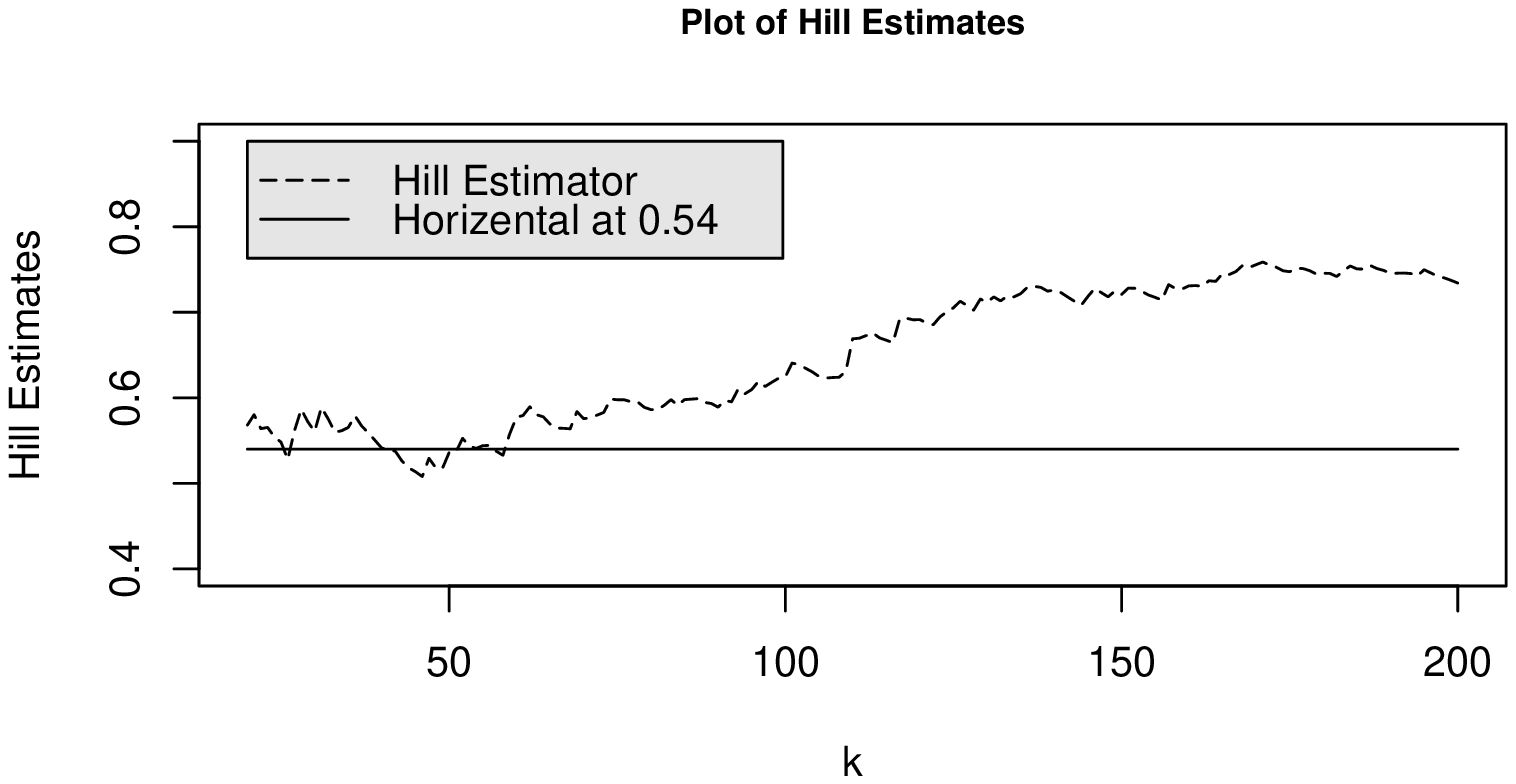}
\end{center}
\vspace{-25pt}
\caption{\sl  Plot of Hill's Estimates}
\label{fig2}
\end{figure}

Since it is difficult to determine a range of $k_n$ such that \eqref{kn} holds,  we employ the Hill plot to help identify a range within which
the Hill estimates are relatively stable. We first provide the plot of Hill's estimates for $\gamma$ for the sample fractions $k_n$ in the range from $10$ to $200$. See Fig.~\ref{fig2}. We observe that a turning point for the Hill estimator occurs around $k_n=58$, and beyond this point the Hill plot shows an obvious upward trend, which indicates that the Hill estimates may be greatly influenced by the biases for large values of $k_n$.  In fact, when $k_n$ is in a range roughly from $26$ to $58$, these Hill estimates are quite close.

\begin{figure}[!htb]
\begin{center}
\includegraphics[scale=0.85]{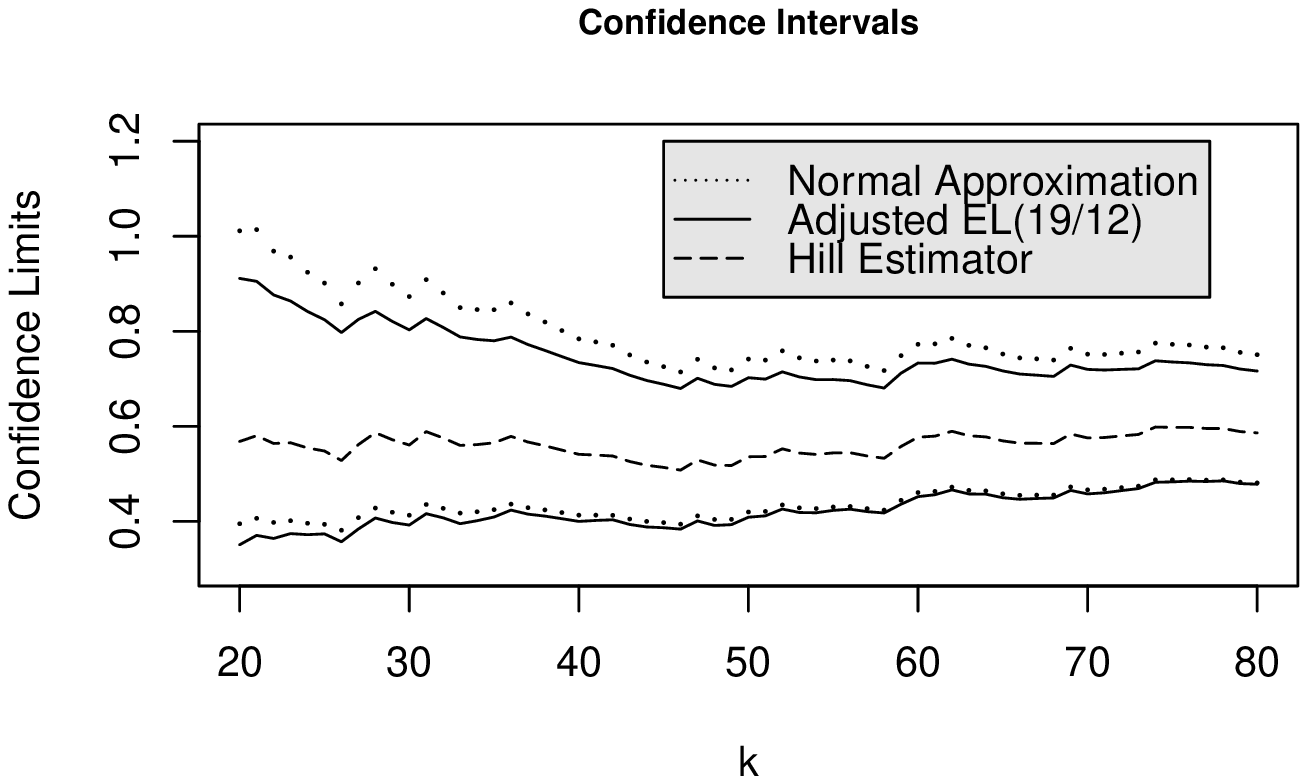}
\end{center}
\vspace{-25pt}
\caption{\sl  Confidence limits:  the dotted lines are upper and lower limits for $95\%$ confidence intervals based on the normal approximation to Hill's estimators, the solid lines are upper and lower limits for $95\%$ confidence intervals based on the adjusted empirical likelihood method with correction factor $a_n=19/12$, and dashed lines are Hill's estimates}
\label{fig3}
\end{figure}

We further compare the upper and lower limits for $95\%$ confidence intervals based on the normal approximation method and adjusted empirical likelihood method. See Fig.~\ref{fig3} for a plot in the range of $k_n$ between $20$ and $80$. These lower confidence limits from two methods are close, but the upper limits of confidence intervals from the adjusted empirical likelihood method are obviously smaller than those from the normal approximation method for Hill's estimator, and the lengths of the confidence intervals from the adjusted empirical method are uniformly shorter than those based on the normal approximation.

\section{Proof of Theorem~\ref{thm1}}\label{proof}

We only provide a sketch of the proof. Note that we use $\gamma_0$ to denote the true value of $\gamma$. Therefore, \eqref{normalapp} is equivalent to
\begin{equation}\label{clt}
\frac1{k_n^{1/2}}\sum^{k_n}_{i=1}y_i(\gamma_0)\xrightarrow{d} N(0,\gamma_0^2).
\end{equation}
From equations (8) and (9) in the proof of Theorem 1 in Lu and Peng (2012), we have
\begin{equation}\label{max}
\max_{1\le i\le k_n}|y_i(\gamma_0)|=o_p(k_n^{1/2}),
\end{equation}
\begin{equation}\label{variance}
\frac{1}{k_n}\sum^{k_n}_{i=1}(y_i(\gamma_0))^2\xrightarrow{p} \gamma_0^2,
\end{equation}

The proof of \eqref{ELchi}, i.e., the proof of Theorem 1 in Lu and Peng (2002), requires only conditions \eqref{clt} to \eqref{variance}.
See Lu and Peng (2002) for details.  Following the same lines in the proof, we only need to verify the following three similar conditions:
\begin{equation}\label{clt1}
\frac1{(k_n+1)^{1/2}}\sum^{k_n+1}_{i=1}y_i(\gamma_0)\xrightarrow{d} N(0,\gamma_0^2),
\end{equation}
\begin{equation}\label{max1}
\max_{1\le i\le k_n+1}|y_i(\gamma_0)|=o_p(k_n^{1/2}),
\end{equation}
\begin{equation}\label{variance1}
\frac{1}{k_n+1}\sum^{k_n+1}_{i=1}(y_i(\gamma_0))^2\xrightarrow{p} \gamma_0^2.
\end{equation}
Since
\[
y_{k_n+1}(\gamma_0)=-\frac{a_n}{k_n}\sum^{k_n}_{i=1}y_i(\gamma_0)=-\frac{a_n}{k_n^{1/2}}\frac{1}{k_n^{1/2}}\sum^{k_n}_{i=1}y_i(\gamma_0)
=O_p(\frac{a_n}{k_n^{1/2}})=o_p(k_n^{1/6}),
\]
equations \eqref{clt1} to \eqref{variance1} follows from \eqref{clt} to \eqref{variance}, respectively.  This completes the proof.

\vspace{10pt}

\noindent{\bf Acknowledgements:}  The authors would like to thank two anonymous referees for their constructive
comments and suggestions that have led to improvements
on the paper.

\end{document}